\documentclass[structabstract]{aa}  
\usepackage{graphicx}
\usepackage{lscape}
\usepackage{natbib}
\bibpunct{(}{)}{;}{a}{}{,}
\usepackage{txfonts}
\begin{document}
   \title{Spectroscopic observations of ices around embedded young stellar objects in the Large Magellanic Cloud with AKARI}


   \author{Takashi Shimonishi
           \inst{1}
           \and
           Takashi Onaka
           \inst{1}
           \and
           Daisuke Kato
           \inst{1}
           \and
           Itsuki Sakon
           \inst{1}
           \and     \\
           Yoshifusa Ita
           \inst{2}
           \and
           Akiko Kawamura
           \inst{3}
           \and
           Hidehiro Kaneda
           \inst{3}           
          }
   \institute{Department of Astronomy, Graduate School of Science, The University of Tokyo, 7-3-1 Hongo, Bunkyo-ku, Tokyo 113-0033, Japan, 
              \email{shimonishi@astron.s.u-tokyo.ac.jp} 
   \and
   National Astronomical Observatory of Japan, 2-21-1 Osawa, Mitaka, Tokyo, 181-8588, Japan
   \and
   Graduate School of Science, Nagoya University, Chikusa-ku, Nagoya, 464-8602, Japan
             }

   \date{}

 
  \abstract
   {}
   {The aim of this study is to understand the chemical conditions of ices around embedded young stellar objects (YSOs) in the metal-poor Large Magellanic Cloud (LMC).}
   {We performed near-infrared (2.5--5$\mu$m) spectroscopic observations toward 12 massive embedded YSOs and their candidates in the LMC using the Infrared Camera (IRC) onboard  {\it AKARI}. 
We estimated the column densities of the H$_2$O, CO$_2$, and CO ices based on their 3.05, 4.27, and 4.67$\mu$m absorption features, 
and we investigated the correlation between ice abundances and physical properties of YSOs.
}
   {The ice absorption features of H$_2$O, CO$_2$, $^{13}$CO$_2$, CO, CH$_3$OH, and possibly XCN are detected in the spectra. 
In addition, hydrogen recombination lines and PAH emission bands are detected toward the majority of the targets. 
The derived typical CO$_2$/H$_2$O ice ratio of our samples ($\sim$0.36 $\pm$ 0.09) is greater than that of Galactic massive YSOs ($\sim$0.17 $\pm$ 0.03), while the CO/H$_2$O ice ratio is comparable. 
It is shown that the  CO$_2$ ice abundance does not correlate with the observed characteristics of YSOs: the strength of hydrogen recombination line and the total luminosity. 
Likewise, clear correlation is not seen between the CO ice abundance and YSO characteristics, but it is suggested that the CO ice abundance of luminous samples is significantly lower than in other samples. 
}
   {The systematic difference in the CO2 ice abundance around the LMC's massive YSOs, which was suggested by previous studies, is confirmed with the new near-IR data. 
We suggest that the strong ultraviolet radiation field and/or the high dust temperature in the LMC are responsible for the observed high abundance of the CO$_2$ ice. 
It is suggested that the internal stellar radiation does not play an important role in the evolution of the CO$_2$ ice around a massive YSO, while more volatile molecules like CO are susceptible to the effect of the stellar radiation. 
}


\titlerunning{Ices around Embedded YSOs in the LMC}

\maketitle


\section{Introduction}   
The infrared spectra of embedded young stellar objects (YSOs) show absorption features originating in various ice species \citep[solid state molecules, e.g.,][]{Ger99, Gib04, Boo08}. 
It is believed that a large amount of heavy elements and complex molecules of the interstellar medium (ISM) are preserved as ices in the dense and cold regions (n$_H$ $\geq$ 10$^4$ cm$^{-3}$, T $\sim$ 10 -- 20K), such as an envelope of a deeply embedded YSO \citep{vDB98, BE04}. 
Since the star formation proceeds in the region of dense molecular clouds, these ices dominate the chemical evolution of YSOs. 
Absorption profiles of ices are known to be sensitive to a chemical composition and a temperature of dust grains, and ices are important tracers of the thermal history of circumstellar environments of YSOs \citep[e.g.,][]{Pon08,Zas09}. 
Thus, investigating the compositions of ices as a function of physical environments is crucial for understanding the chemical evolution of YSOs and is one of the key topics in the current astrochemistry. 
Ices are also detected toward solar system objects such as comets, icy satellites, and Mars. 
Interstellar ices are thought to be taken into planets and comets as a result of subsequent planetary formation activities \citep{Ehr00}. 
Therefore the chemical evolution of ices around YSOs is also of interest in terms of chemical conditions of the planetary formation.

 Observations of ices around extragalactic YSOs are one of the challenges in current ice studies \citep{ST,Oli09,Sea09,vanL05,vanL10}. 
So far, infrared spectroscopic observations of extragalactic embedded YSOs are still limited by observational difficulties, and their circumstellar chemistry is poorly understood. 
However, it is  probable that different galactic environments (e.g., metallicity or radiation field) could affect the properties of circumstellar material.  Thus detailed investigations of extragalactic YSOs provide us important information for understanding the diversity of chemical conditions of YSOs in the universe. 
Thanks to the progress of infrared space telescopes, such as {\it AKARI} \citep{Mur07} and {\it Spitzer} \citep{Wer04}, we are now able to extend the study of ices around YSOs to extragalactic objects.

 The Large Magellanic Cloud (LMC), the nearest irregular galaxy to our Galaxy, offers an ideal environment for studying extragalactic star formations since individual YSOs can be identified with reasonable spatial resolution (1"$\sim$0.25pc) thanks to its proximity \citep[$\sim$50kpc;][]{Alv04}. 
A unique metal-poor environment of the LMC \citep[Z $\sim$ 0.3Z$_{\odot}$,][]{Luc98} results in a generally high UV radiation field over the galaxy, and these factors can affect the chemical conditions of circumstellar materials. 
The nearly face-on viewing angle of this galaxy \citep[$\sim$35 degrees,][]{MC01} allows correlation studies of individual YSOs and the ISM. 
Because of these advantages, various types of surveys have been performed toward the LMC, which provide information of both ISM and stars over a wide wavelength range \citep[e.g.,][]{Zar04,Mei06,Kat07}.

 \citet{ST} spectroscopically confirmed seven massive YSOs in the LMC based on the {\it AKARI} LMC spectroscopic survey \citep{Ita08} and reported the detection of the H$_2$O and CO$_2$ ices toward these objects. 
\citet{Oli09} analyzed the 15.2$\mu$m feature of the CO$_2$ ice for 15 embedded YSOs in the LMC and discuss the chemical properties of the ice mantles. 
The H$_2$O and CO$_2$ ices are ubiquitous and are the major components of interstellar ices. 
Based on the low-resolution (R $\sim$ 20) spectra obtained in \citet{ST}, we derived a typical CO$_2$/H$_2$O ice ratio and showed that the ratio around the LMC's massive YSOs (0.45$\pm$0.17) is higher than previously reported toward Galactic massive YSOs \citep[0.17$\pm$0.03][]{Ger99}. 
The result suggests the chemically different nature of YSOs in the metal-poor galaxy.

 An accurate determination of column densities of major ice species is crucial when discussing the variation of ice abundances between the objects.
However, the uncertainties in the derived column densities from our previous low-resolution spectra are large, and the number of samples for which the column densities of major ice species are determined is still small. 
In addition, the features of relatively minor ice species, such as CH$_3$OH and CO are difficult to identify by the low-resolution spectra. 
Investigation of these minor ices should help in constraining the processing of ices in a YSO envelope \citep[e.g.,][]{Dar99,Thi06}. 
Thus follow-up observations with higher spectral resolution are required to further understand the chemical conditions of extragalactic YSOs.

 In the present study, we used the higher spectral resolution (R $\sim$ 80) mode of the Infrared Camera \citep[IRC,][]{TON07} onboard {\it AKARI} to observe massive embedded YSOs and their candidates in the LMC. 
The correlation between the chemical properties of ices and the YSO characteristics, which could not be discussed in \citet{ST} due to the large uncertainties of the observed spectra, is discussed in this paper. 
In addition, the improved calculation method of the ice column density by using the curve-of-growth is presented in an Appendix.



\section{Targets}  

 The sample of massive YSOs presented in this study is mainly selected from \citet{ST}, which spectroscopically discovered several massive YSOs in the LMC by using a combination of the low-resolution, near-infrared spectroscopic survey and the color-color diagram. 
The H$_2$O and CO$_2$ ice features toward these targets have already been reported in \citet{ST}, and this indicates the embedded nature of these YSOs. 
In addition, we include several YSO ``candidates''. 
These objects were selected by the same photometric criteria described in \citet{ST} and their infrared spectral energy distributions (SEDs) are red enough to be identified as a YSO. 
But absorption features of ices seen in their low-resolution 2--5$\mu$m spectra are too weak to definitely classify them as YSOs, so they are not included in the YSO list of \citet{ST}. 
Some other targets were selected based on the YSO candidates catalog presented in \citet{Whin08}, which performed photometric selection of YSOs using the database of {\it Spitzer} SAGE project \citep{Mei06}. 
We selected relatively bright and red objects from their catalog. 
Van Loon et al. (2005) observed IRAS05328-6827 with the Infrared Spectrograph \citep[IRS,][]{Hou04} onboard {\it Spitzer} and the Infrared Spectrometer And Array Camera (ISAAC) at the Very Large Telescope (VLT), and report the presence of ice features in its near- to mid-infrared spectrum. 
This object is also added to the present targets.

 To understand the nature of the targets, we carried out the SED fitting of each YSO by using the Online SED Fitter\footnote{The Online SED Fitter is available at http://caravan.astro.wisc.edu/protostars/sedfitter.php} \citep{Rob07}. 
The photometric dataset used for the SED fitting are obtained from the available database and recent YSO catalogs \citep{Zar04,Mei06,Kat07,Ita08, GC09}, 
and 2--24$\mu$m data were typically used as the input data. 
The estimated total luminosities and masses of our targets are $\sim$4--340$\times$10$^3L_{\odot}$ and 12--47$M_{\odot}$, respectively. 
Thus the targets in the present study are massive YSOs. 
Details of all the targets are listed in Table \ref{tbl_NG}.


\begin{table*}[!]
\caption{Targets of the present observations}
\label{tbl_NG}
\centering
\renewcommand{\footnoterule}{}  
\begin{tabular}{ l c c c c c l }
\hline\hline
Number & ID                                      & Obs. ID &  Obs. Date  &  RA[J2000]      & DEC[J2000]    & Reference  \\
\hline
ST1    & IRAS05400-7013	                         & 	 5126560    &  2008 Apr 12 &  05:39:31.15 	 & -70:12:16.8 	 & 1, 2	   \\	   
ST2    & NGC 1936                          	     & 	 5126562    &  2008 May 25 &  05:22:12.56 	 & -67:58:32.2 	 & 1	   \\	   
ST3    & ......$^a$                              & 	 3410003    &  2008 Dec 11 &  05:25:46.69 	 & -66:14:11.3 	 & 1, 3	   \\	   
ST4    & IRAS F05148-6715                  	     & 	 3410004    &  2008 Dec 27 &  05:14:49.41 	 & -67:12:21.5 	 & 1, 3	   \\	   
ST5    & IRAS 05311-6836                   	     & 	 5126561    &  2008 May 11 &  05:30:54.27 	 & -68:34:28.2 	 & 1, 2	   \\	   
ST6    & 05394112-6929166$^b$            	     & 	 5126563    &  2008 Apr 19 &  05:39:41.08 	 & -69:29:16.8 	 & 1	   \\	   
ST7    & IRAS05240-6809                    	     & 	 5126567    &  2008 May 23 &  05:23:51.15 	 & -68:07:12.2 	 & 1, 3	   \\	   
ST8    & MSX LMC 1786                      	     & 	 5126573    &  2008 May  2 &  05:37:28.17 	 & -69:08:47.0 	 & 3, 4, 5, 6  \\	   
ST9    & IRAS05237-6755                          & 	 5126568    &  2008 May 26 &  05:23:35.53 	 & -67:52:35.5 	 & 2, 4    \\	   
ST10   & MSX LMC 1229                      	     & 	 5200384    &  2008 Dec 21 &  04:56:40.80 	 & -66:32:30.4 	 & 2, 4, 5 \\	   
ST11   & IRAS05270-6851	                         &   5126571    &  2008 May 11 &  05:26:46.63	 & -68:48:47.1	 & 2, 5    \\	   
ST12   & IRAS05328-6827                   	     & 	 5126566    &  2008 May 12 &  05:32:38.59 	 & -68:25:22.2 	 & 3, 6	   \\      
\hline
\end{tabular}
\begin{flushleft}
$^a$The source is in a cluster,  $^b$2MASS ID \\
\textbf{References.} (1) \citet{ST}, (2) \citet{Sea09}, (3) \citet{Oli09}, (4) \citet{Whin08}, (5) \citet{GC09}, (6) \citet{vanL10}
\end{flushleft}
\end{table*}


\section{Observation and data reduction}  
 Spectra of 2--5$\mu$m presented in this paper were obtained as a part of {\it AKARI} post-helium open time program ``Ices Around Extragalactic Young Stellar Objects'' (IEYSO, PI: T. Shimonishi) and {\it AKARI} Director's Time (DT) observations. 
{\it AKARI} is the first Japanese satellite dedicated to infrared astronomy launched in 2006. 
The telescope system and the scientific instruments onboard were cooled down by the liquid helium with mechanical coolers. 
The {\it AKARI} cryogen boiled off on 2007 August 26, 550 days after launch. 
All of the data presented here were obtained after the exhaustion of liquid helium.

The slit-less NG spectroscopy mode with the Np point source aperture was used to obtain spectra between 2.5 and 5$\mu$m with a spectral resolution of R$\sim$80. 
The wavelength accuracy of the spectroscopy is estimated to be $\sim$0.01--0.015$\mu$m. 
The spatial resolution of a point source is approximately 5--8 arcseconds. 
Overlapping of dispersed spectra with other objects is a serious problem for slit-less spectroscopies, which makes extraction of reliable spectra difficult. 
The use of the Np point source aperture reduces the effect of spectral overlapping and enables  spectra to be extracted from the objects located in crowded regions \citep[see ][and {\it AKARI} Data Users Manual for details of the IRC spectroscopy]{Ohy07}. 

 The basic spectral analysis was performed using the standard IDL pipeline prepared for reducing {\it AKARI} IRC post-helium mission data. 
Raw data were converted to dark-subtracted and linearity-corrected frames in the pipeline. 
We used a newly calibrated spectral response curve to analyze the present spectra. 
At that time, appropriate flat field data were not available for post-helium Np data. 
However, because of the faintness of the natural background and the small pixel-to-pixel variation of the flux at the Np aperture, there are no significant differences between the spectra with and without flat-fielding ({\it AKARI} Data Users Manual ver. 1.3). 
In fact, spectra processed without the flat-fielding is better in terms of S/N because of the relatively low S/N of the flat field data. 
Therefore, we did not perform the flat-fielding in the pipeline. 
Four to seven pixels were integrated in the spatial direction to extract the point source spectrum. 
Each pixel was smoothed in the dispersion direction across the width of three pixels to suppress the effect of bad pixels. 
Background signals of each source were estimated from the signal counts of the adjacent region of the target spectrum and subtracted.


\section{Results}  

\subsection{Observed spectra}  

\begin{table*}
\caption{Details of detected ice features}
\label{ice_detail}
\centering
\renewcommand{\footnoterule}{}  
\begin{tabular}{ c c c c c }
\hline\hline
Molecule       &  Wavelength &   Vibration Mode     &   A$^{a}$                        &  Reference               \\
     ~         &  [$\mu$m]   &                      &  [10$^{-17}$cm molecule$^{-1}$]  &     ~                    \\
\hline
H$_2$O         &  3.05       &  O-H stretch         &  20                              &  1                     \\
CH$_3$OH       &  3.53       &  C-H stretch         &  0.76                            &  2                     \\
CO$_2$         &  4.27       &  C-O stretch         &  7.6                             &  1                     \\
$^{13}$CO$_2$  &  4.38       &  $^{13}$C-O stretch  &  7.8                             &  1                     \\
XCN            &  4.62       &  CN stretch          &  ...                             &  3                     \\
CO             &  4.67       &  C-O stretch         &  1.1                             &  1                     \\
\hline 
\end{tabular}
\begin{flushleft}
$^a$Band strengths of absorption features                                   \\
\textbf{References.} (1) \citet{Ger95}, (2) \citet{dHA86}, (3) \citet{Sch97} \\
\end{flushleft}
\end{table*}

\begin{figure*}[!htb]
\begin{center}
\includegraphics[width=15cm, angle=0]{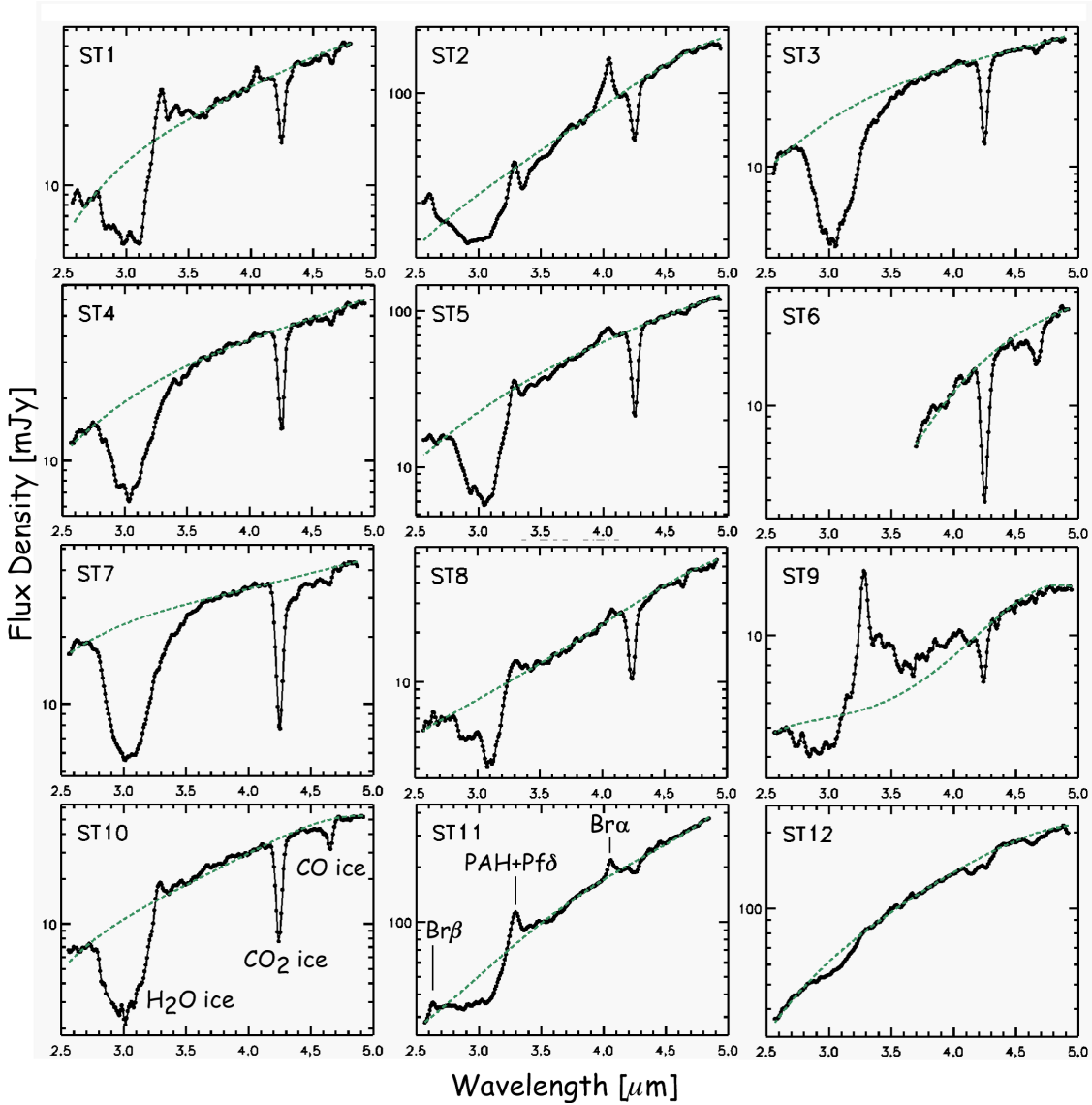}
\caption{{\it AKARI} IRC NG 2.5--5$\mu$m spectra of YSOs in the LMC. 
Dashed lines represent derived continuum. 
The positions of detected features are labeled.}
\label{NG_rsults}
\end{center}
\end{figure*}

 Observed spectra for all the objects are shown in Fig.~\ref{NG_rsults} and details of the detected ice features are summarized in Tables \ref{ice_detail} and \ref{ice_detect}. 
The absorption feature of the 3.05$\mu$m H$_2$O ice is detected toward all the objects except ST6 and ST9. 
The profile of the H$_2$O ice feature is resolved, and this permits more accurate determination of the column density than in \citet{ST}. 
One object (ST8) shows a structured profile of the H$_2$O ice, which is often linked to the thermal processing of the ice mantles \citep{Obe07}. 
The water-ice band of ST6 is seen in the spectrum of \citet{ST}, but it is not detected in the present spectra owing to the slightly poor sensitivity of the present observation mode. 
The 3$\mu$m region of ST9 overlaps prominent emission components, which makes it difficult to identify the underlying absorption bands around 3$\mu$m. 
A broad depression is seen in the 4.3 -- 4.7$\mu$m region of ST7. 
It can probably be attributed to a combination mode of the H$_2$O ice, which typically peaks around 4.5$\mu$m \citep{Gib00}. 


\begin{table}
\caption{Detection of ice features in our YSO spectra}
\label{ice_detect}
\centering
\renewcommand{\footnoterule}{}  
\begin{tabular}{ c c c c c c c }
\hline\hline

Number  &   H$_2$O  &  CO$_2$   &    CO     &  CH$_3$OH & $^{13}$CO$_2$ & XCN \\
\hline
ST1     &  $\surd$  &  $\surd$  &  $\surd$  &    ---    &    ---    &    ---  \\
ST2     &  $\surd$  &  $\surd$  &    ---    &  $\surd$  &     ?     &    ---  \\
ST3     &  $\surd$  &  $\surd$  &  $\surd$  &    ---    &    ---    &    ---  \\
ST4     &  $\surd$  &  $\surd$  &  $\surd$  &    ---    &    ---    &    ---  \\
ST5     &  $\surd$  &  $\surd$  &  $\surd$  &  $\surd$  &    ---    &    ---  \\
ST6     &    ---    &  $\surd$  &  $\surd$  &    ---    &    ---    &     ?   \\
ST7     &  $\surd$  &  $\surd$  &  $\surd$  &    ---    &  $\surd$  &     ?   \\
ST8     &  $\surd$  &  $\surd$  &  $\surd$  &    ---    &  $\surd$  &    ---  \\
ST9     &     ?     &  $\surd$  &  $\surd$  &    ---    &     ?     &    ---  \\
ST10    &  $\surd$  &  $\surd$  &  $\surd$  &    ---    &    ---    &     ?   \\
ST11    &  $\surd$  &  $\surd$  &    ---    &     ?     &    ---    &    ---  \\
ST12    &  $\surd$  &  $\surd$  &    ---    &     ?     &    ---    &    ---  \\
\hline 
\end{tabular}
\begin{flushleft}
\textbf{Note.} $\surd$ = clear detection, ? = tentative detection, --- = no detection. \\
\end{flushleft}
\end{table}

The absorption feature of the 4.27$\mu$m CO$_2$ ice is detected toward all the objects. 
Some objects show the absorption features of 3.53$\mu$m CH$_3$OH ice, 4.38$\mu$m $^{13}$CO$_2$ ice, 4.62$\mu$m XCN, and the 4.67$\mu$m CO ice. 
Since the CO$_2$ and CO ice features are not completely resolved in our spectra, the widths of the observed features are larger than those of their typical laboratory spectra. 
The detections of the $^{13}$CO$_2$ ice and XCN features are still tentative because of the unresolved absorption bands of gaseous CO, which sometimes appear in the wavelength region of these features \citep{Chi98,Boo00}. 
The non-detection of the $^{13}$CO$_2$ ice in ST10, on the other hand, gives a rough lower limit of $^{13}$CO$_2$/$^{12}$CO$_2$ of about 95. 
This is in the same range as the isotopic ratio for Galactic YSOs \citep{Boo00}. 
These minor features are labeled in Fig.~\ref{ice_results1} and summarized in Table \ref{ice_detect}. 

\begin{figure*}[!ht]
\begin{center}
\includegraphics[width=15.0cm, angle=0]{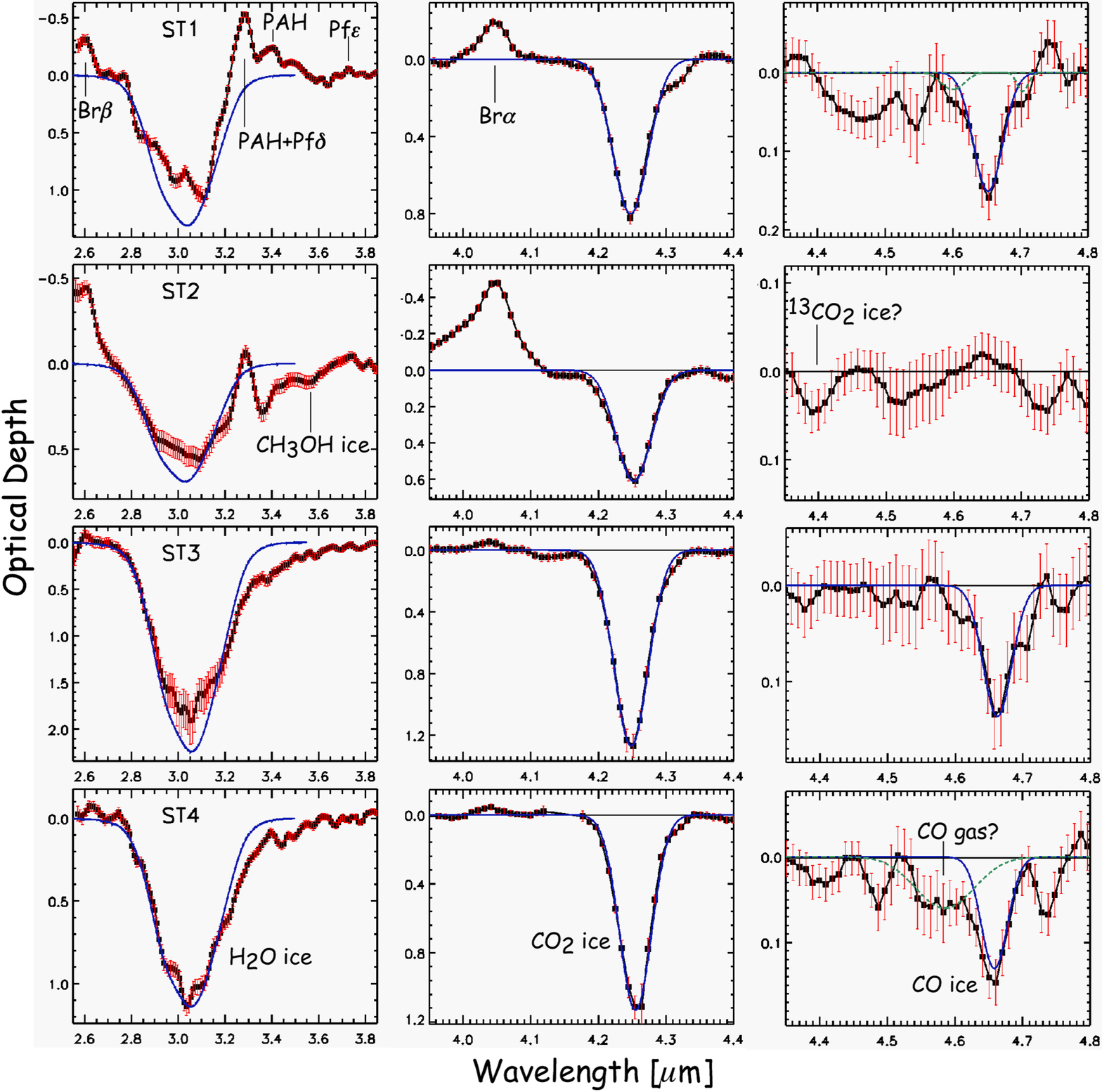}
\caption{Observed spectra of sample YSOs after the continuumd subtraction. 
A vertical axis represents the optical depth. 
The positions of the detected features are labeled.
(Left) 2.6 -- 3.8$\mu$m region. 
A solid line represents the fitted laboratory spectrum of the H$_2$O ice \citep[pure H$_2$O at 10K,][]{Ehr96}. 
2.9 -- 3.1$\mu$m (bottom of the H$_2$O ice feature) and 3.2 -- 3.4$\mu$m are removed from the fitting wavelength. 
(Center) 3.9 -- 4.4$\mu$m region. 
A solid line represents a fitted Gaussian profile for a measurement of the equivalent width of the CO$_2$ ice feature.  
(Right) 4.4 -- 4.8$\mu$m region. 
A dashed line indicates unresolved CO gas bands and the possible XCN feature, which were subtracted when fitting the CO ice feature (solid line). 
}
\label{ice_results1}
\end{center}
\end{figure*}

\setcounter{figure}{1}
\begin{figure*}[!ht]
\begin{center}
\includegraphics[width=15cm, angle=0]{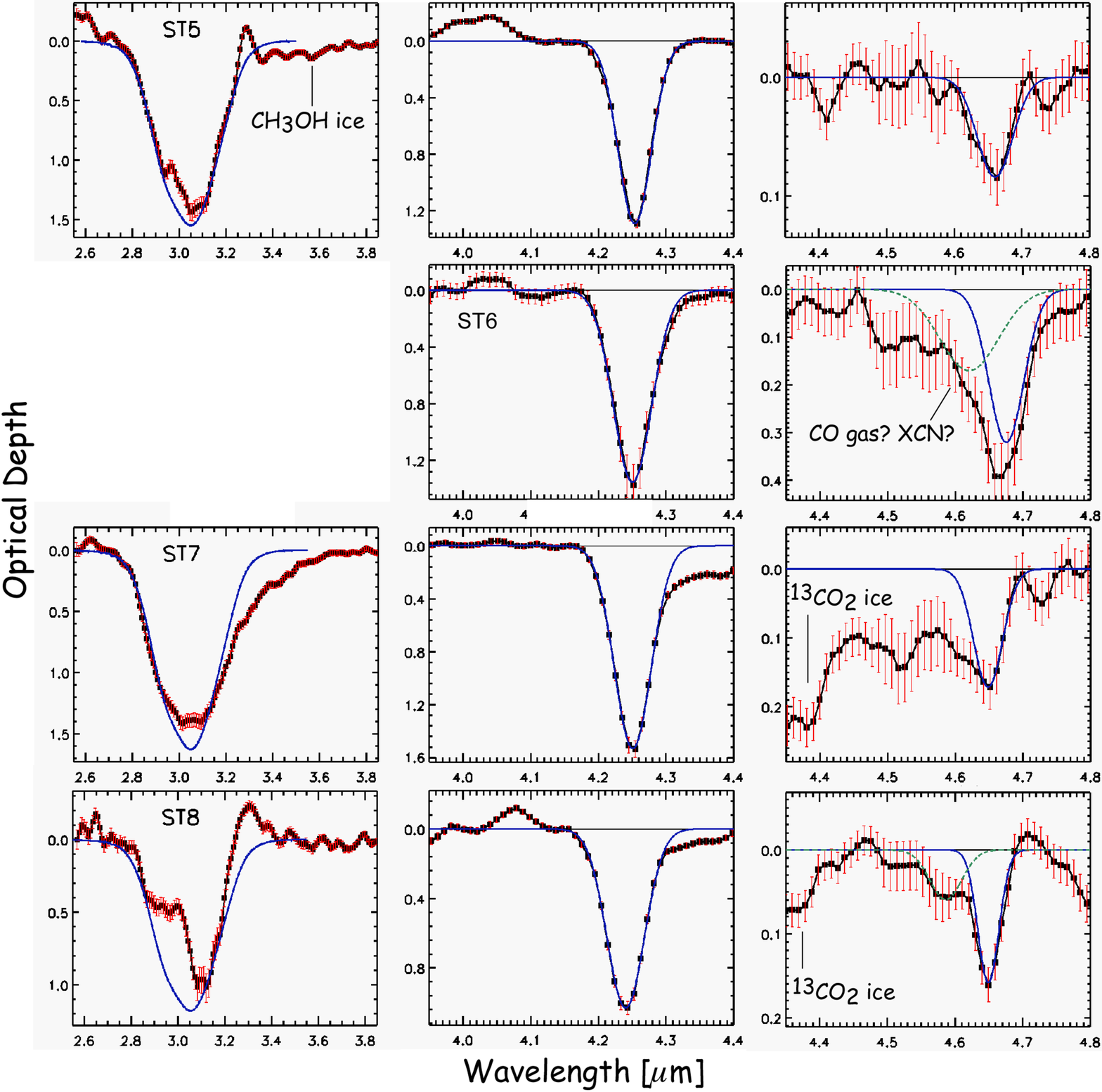}
\caption{\textit{Continued}}
\end{center}
\label{ice_results2}
\end{figure*}

\setcounter{figure}{1}
\begin{figure*}[!ht]
\begin{center}
\includegraphics[width=15cm, angle=0]{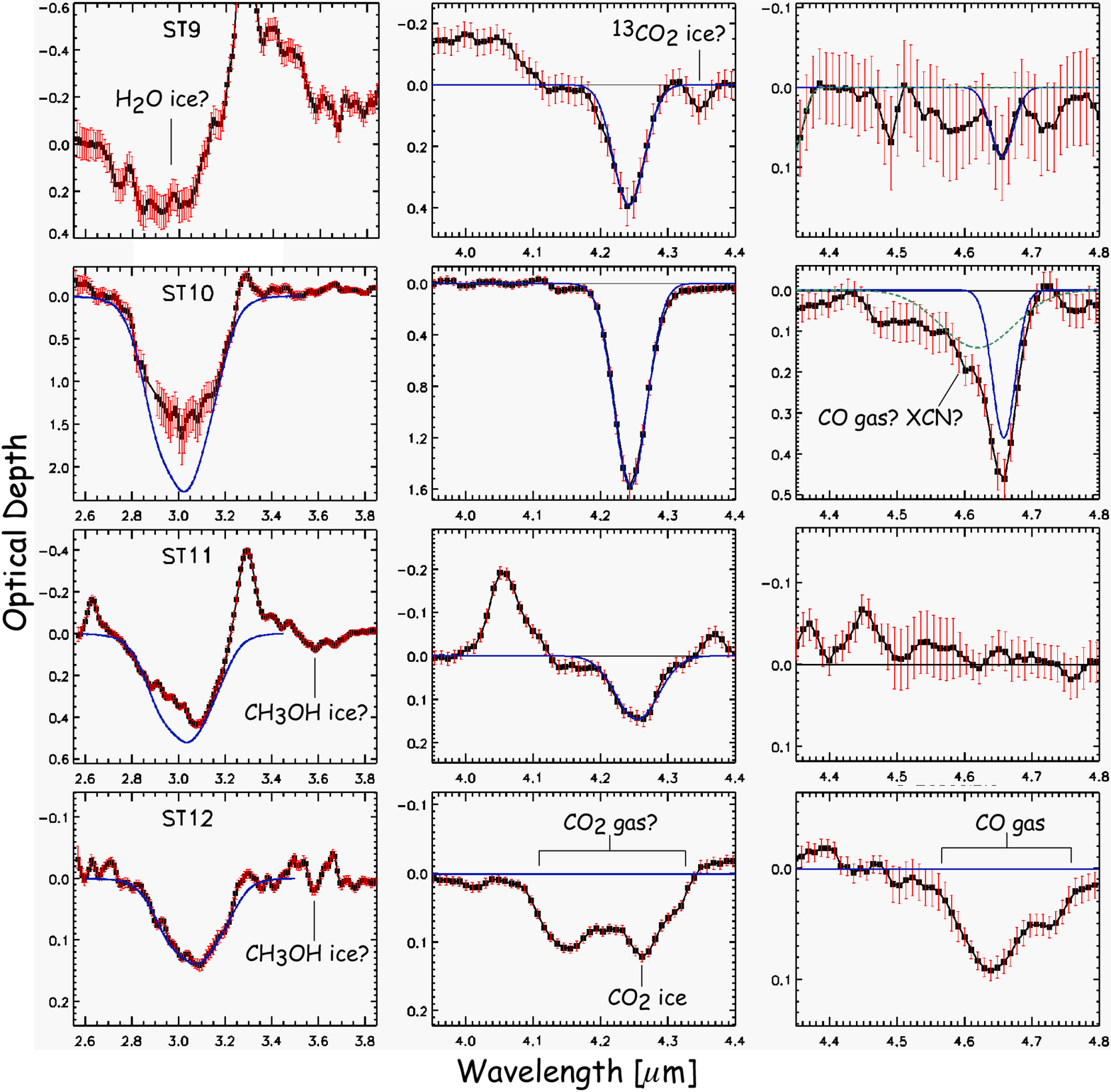}
\caption{\textit{Continued}}
\end{center}
\label{ice_results3}
\end{figure*}

Several emission features that were not resolved in the low-resolution spectra of \citet{ST} are detected in the present spectra. 
They are hydrogen recombination lines (2.62$\mu$m Br$\beta$, 3.05$\mu$m Pf$\epsilon$, 3.29$\mu$m Pf$\delta$, 3.74$\mu$m Pf$\gamma$, 4.05$\mu$m Br$\alpha$, 4.64$\mu$m Pf$\beta$) and polycyclic aromatic hydrocarbon(PAH) emission bands (3.3$\mu$m and 3.4$\mu$m). 
Detection of these emission features allows detailed discussions about the correlation between ice abundances and YSO characteristics (see $\S$5.2).


\subsection{Spectral fitting and calculation of ice column densities}  

 To subtract the continuum, we fitted a polynomial of the second to fourth order to the continuum regions and derived the optical depth spectra. 
The wavelength regions used for the continuum were set to avoid a region with emission or absorption features, and they were typically 2.5--2.7 $\mu$m, 4.1--4.15 $\mu$m, and 4.85--4.95 $\mu$m. 
The absorption features of ices except H$_2$O were not completely resolved with our spectral resolution. 
Therefore, to calculate ice column densities, we used different methods for the H$_2$O ice and the other ices.

\subsubsection{H$_2$O ice} 
 The spectral resolution of the present spectra is sufficient to resolve a broad feature of the H$_2$O ice, and the true optical depth can be measured. 
Thus a direct comparison of laboratory spectra is possible. 
We fitted laboratory ice profiles to the observed spectra by a $\chi^2$ minimization method and derived the column density from the equation
\begin{equation}
N = \int \tau  d\nu / A, 
\end{equation}
where $A$ is the band strength of each ice feature (in units of cm molecule$^{-1}$) as measured in the laboratory, $N$ the column density in cm$^{-2}$, $\nu$ the wavenumber in units of cm$^{-1}$, and $\tau$ the optical depth. 
The integration is performed over the H$_2$O ice absorption band. 
The wavelength range used for the fit is set to avoid the 3.3$\mu$m PAH emission band and the bottom of the H$_2$O absorption feature, which typically suffers from the weak background emission. 
Typically, 2.7--2.9$\mu$m and 3.1--3.2$\mu$m are used for the fitting. 
A single laboratory ice profile of the pure H$_2$O ice at 10K, which is expected to exist in the cold envelope of YSOs, is fitted to the observed spectra
The laboratory spectra of the H$_2$O ice was taken from the Leiden Molecular Astrophysics database \footnote{data available at http://www.strw.leidenuniv.nl/~lab} \citep{Ehr96}. 
We adopted the band strengths of the H$_2$O ice band as 2.0$\times$10$^{-16}$ cm molecule$^{-1}$ \citep{Ger95}. 
The derived column densities are listed in Table \ref{tbl_NG_ice}, and results of the fitting are shown in Fig.~\ref{ice_results1}. 

\begin{table*}[!ht]
\caption{Column density of ices and properties of YSOs}
\label{tbl_NG_ice}
\centering
\renewcommand{\footnoterule}{}  
\begin{tabular}{ l c c c c c c c }
\hline\hline
Number & N(H$_2$O)                 & N(CO$_2$)               & N(CO$_2$)/N(H$_2$O) & N(CO)                  & N(CO)/N(H$_2$O) & EW(Br$\alpha$) &  L$_{tot}$ \\
   ~   & [10$^{17}$ cm$^{-2}$]     & [10$^{17}$ cm$^{-2}$]   & [$\%$]              & [10$^{17}$ cm$^{-2}$]  & [$\%$]              & [$\AA$]        &  [10$^3$$\times$L$_{\odot}$]   \\
\hline
ST1    & 22.5 $\pm$ 0.96           &   5.66 $\pm$ 0.64       & 25                  & 3.62 $\pm$ 0.93        & 16   & 100 $\pm$ 7     &  53 $\pm$ 18          \\
ST2    & 11.9 $\pm$ 1.0            &   3.64 $\pm$ 0.38       & 31                  & $<$0.5                 & $<$1 & 343 $\pm$ 12    &  182 $\pm$ 9          \\
ST3    & 38.1 $\pm$ 3.9            &   10.6 $\pm$ 1.6        & 28                  & 3.17 $\pm$ 0.92        & 8    & 21.3 $\pm$ 6.3  &  9.9 $\pm$ 1.0        \\
ST4    & 19.6 $\pm$ 0.83           &   7.41 $\pm$ 0.79       & 38                  & 2.71 $\pm$ 1.2         & 14   & 17.4 $\pm$ 5.1  &  4.0 $\pm$ 0.4        \\
ST5    & 26.4 $\pm$ 1.2            &   10.1 $\pm$ 0.79       & 39                  & 2.33 $\pm$ 0.82        & 9    & 119 $\pm$ 7     &  66 $\pm$ 12          \\
ST6    & 45.6 $\pm$ 16$^a$         &   17.1 $\pm$ 3.0        & 37                  & $<$13                  & ...  & 44.3 $\pm$ 21   &  6.6 $\pm$ 2.5        \\
ST7    & 27.8 $\pm$ 0.95           &   19.5 $\pm$ 1.8        & 70                  & $<$4                   & $<$14 & 15.8 $\pm$ 6.1 &  20 $\pm$ 5           \\
ST8    & 20.1 $\pm$ 1.2            &   9.34 $\pm$ 1.1        & 47                  & 2.60 $\pm$ 0.75        & 13   & 64.1 $\pm$ 6.2$^c$  &  46 $\pm$ 16$^d$  \\
ST9    & ...                       &   1.84 $\pm$ 0.17       & ...                 & $<$2                   & ...  & ...             &  54 $\pm$ 22          \\
ST10   & 39.8 $\pm$ 5.0            &   14.5 $\pm$ 2.1        & 36                  & 7.72 $\pm$ 1.5         & 19   & $<$10           &  32 $\pm$ 14          \\
ST11   & 8.9 $\pm$ 0.3             &   0.9 $\pm$ 0.23        & 10                  & $<$0.5                 & $<$1 & 116 $\pm$ 6.    &  338 $\pm$ 27         \\
ST12   & 2.34 $\pm$ 0.12           &   0.9 $\pm$ 0.15$^b$    & 39                  & ...                    & ...  & $<$7            &  23 $\pm$ 3$^d$       \\
\hline
\end{tabular}
\begin{flushleft}
$^a$Derived from low-resolution spectrum of \citet{ST}, but re-calculated after applying our own spectral reduction pipeline   \\
$^b$Derived by \citet{Oli09}   \\
$^c$Detection of a strong [OIII] emission line at 88 $\mu$m is reported by \citet{vanL10}.  \\
$^d$Estimated luminosities 95 and 62L$_{\odot}$ from \citet{vanL10}, on the basis of far-infrared data, for ST8 and ST10, respectively. 
\end{flushleft}
\end{table*}

\subsubsection{CO$_2$ ice and CO ice}  
 In case the spectral resolution is insufficient to resolve the absorption features, we cannot derive the ``true'' optical depth from the spectra and we can not use Eq. (1) to calculate column densities. 
Instead, we use the equivalent width of the absorption features to calculate column densities since it does not depend on the spectral resolution. 
The equivalent width ($EW$) is defined as 
\begin{equation}
EW = \int F_{norm} d\lambda, 
\end{equation}
\begin{equation}
\textrm{where} \; F_{norm} = 1- F_{obs} / F_c, 
\end{equation}
and $F_{norm}$ is the normalized flux, $F_{obs}$ and $F_c$ are the observed and continuum flux, and $\lambda$ is the wavelength. 
A Gaussian profile with the fixed central wavelength was fitted to the CO$_2$ and CO absorption features to measure the equivalent width. 
To convert the equivalent width to the column density, we need to calculate the ``curve-of-growth'' of each absorption feature. 
The detailed discussion about our method is described in Appendix.A. 
To calculate the curve-of-growth, we need to assume an intrinsic profile of the target absorption band. 
It is known that the absorption profiles of the CO$_2$ and CO ices change depending on the temperature and compositions of the ices \citep{Ger99,Chi98}. 
By considering the observed variation in ice profiles around massive YSOs, we choose appropriate profiles for the CO$_2$ and CO ice and estimated the error of our analysis (see App. A for details). 
We adopted the band strengths of the CO$_2$ and CO ices to be 7.6$\times$10$^{-17}$ and 1.1$\times$10$^{-17}$ cm molecule$^{-1}$, respectively \citep{Ger95}.

Gas-phase CO$_2$ shows narrow absorption lines around the wavelength range of the CO$_2$ ice feature, and it is difficult to resolve it with our spectral resolution. 
However, the contribution of the gas-phase CO$_2$ is generally small toward YSOs \citep{Ger99,Num01}, and we assume that it has a negligible effect on the derived equivalent widths, except for ST12, which shows a strong absorption band of the gas-phase CO$_2$. 
On the other hand, the gas-phase CO ro-vibrational lines sometimes appear in the spectra of YSOs and are blended with the CO ice feature with our spectral resolution \citep{Chi98}. 
Unresolved gas-phase CO absorption bands are expected to cause shallow and broad absorption features on either or both sides of the CO ice feature. 
Therefore, in cases where these bands are seen on the sides of the CO ice feature, we subtracted them by assuming a Gaussian profile. 
The derived column densities are listed in Table \ref{tbl_NG_ice}.

\subsubsection{CH$_3$OH ice, $^{13}$CO$_2$ ice, and XCN} 
 In this study, we did not estimate the column densities of these minor ice species because the observed strengths of these ice features are too weak to discuss the column densities with the present S/N. 
In addition, it is difficult to distinguish the $^{13}$CO$_2$ ice and XCN features from the unresolved CO gas absorption bands by our spectral resolution. 
Observations of these minor ice features with higher spectral resolution and better S/N are challenges for the future.


\section{Discussion}  

\subsection{Large CO$_2$/H$_2$O ice ratio in the LMC}

\begin{figure}[!htb]
\includegraphics[width=\hsize, angle=0]{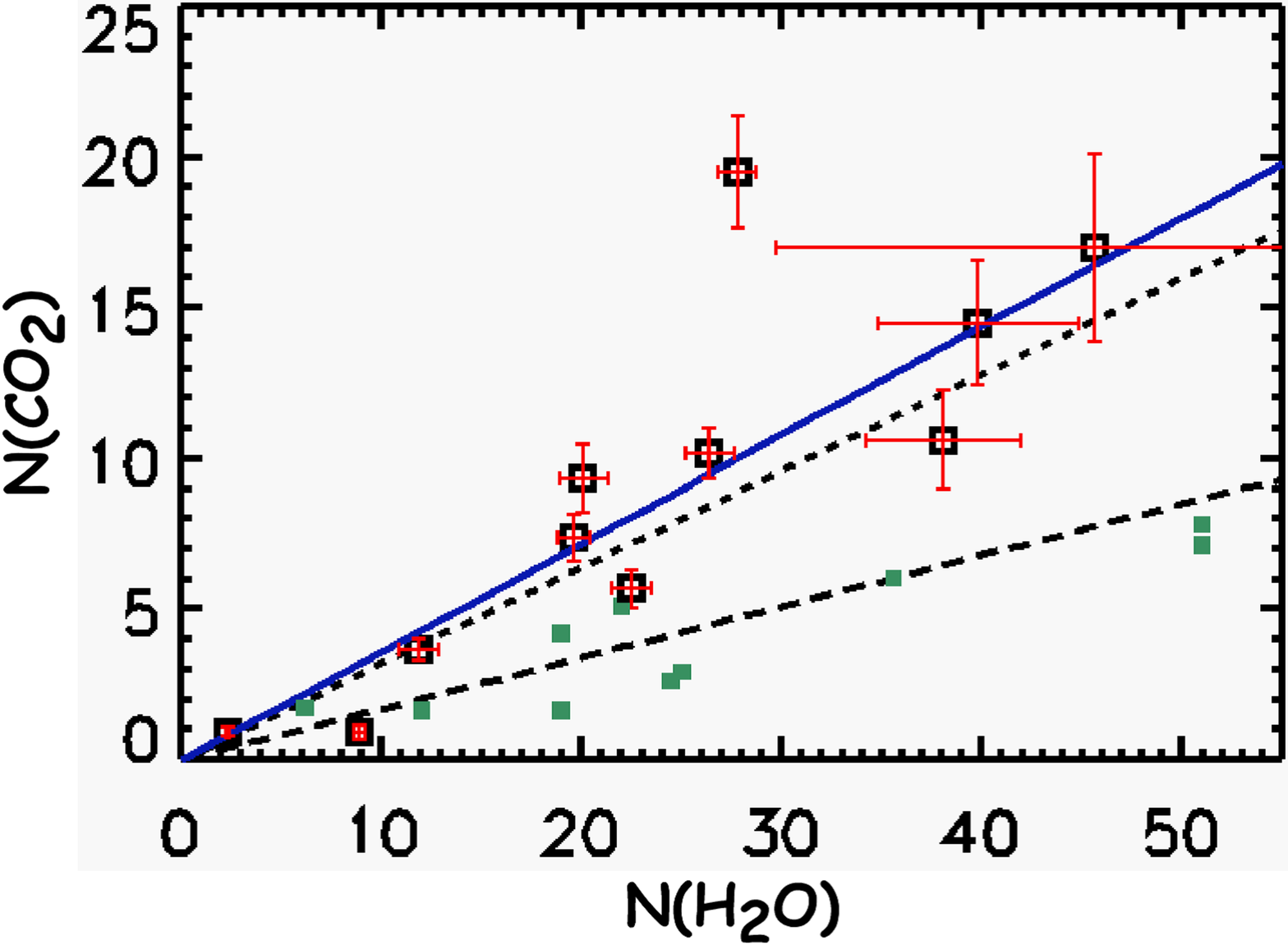}
\caption{
CO$_2$ ice vs. H$_2$O ice column density in units of 10$^{17}$ cm$^{-2}$. 
Open squares with error bars represent the results of this study. 
Filled squares represent those of Galactic massive YSOs \citep{Ger99, Gib04}. 
An average CO$_2$/H$_2$O ice ratio of our samples (0.36) is plotted as a solid line. 
Dotted and dashed lines represent CO$_2$/H$_2$O ratio $\sim$ 0.32 and 0.17, which correspond to the typical CO$_2$/H$_2$O ratio of Galactic low- and intermediate-mass YSOs, and massive YSOs, respectively. 
The dotted line also corresponds to the CO$_2$/H$_2$O ratio of the LMC's massive YSOs derived by \citet{Oli09}. 
}
\label{NG_H2O_CO2}
\end{figure}

The derived column densities of the H$_2$O and CO$_2$ ices are plotted in Figure \ref{NG_H2O_CO2}. 
The average CO$_2$/H$_2$O ice column density ratio of our samples is calculated to be 0.36 $\pm$ 0.09. 
The present samples are massive YSOs and should be compared to Galactic massive samples. 
Thus the column densities of Galactic massive YSOs \citep[CO$_2$/H$_2$O $\sim$ 0.17 $\pm$ 0.03,][]{Ger99,Gib04} are also plotted in the figure for comparison. 
It is reported that Galactic quiescent molecular clouds show a similar CO$_2$ ice abundance \citep[$\sim$ 0.18 $\pm$ 0.04,][]{Whit07}. 
As can be seen in the figure, the typical abundance\footnote{The ``abundance'' of a given ice is generally defined as the ratio of the column density relative to the H$_2$O ice since it is the most abundant ice species.} of the CO$_2$ ice toward the LMC's massive YSOs is larger than for Galactic massive YSOs. 
Also, the scatter of the CO$_2$/H$_2$O ratio is larger than for Galactic samples. 

\citet{ST} reported that the typical CO$_2$/H$_2$O ice ratio in the LMC is 0.45 $\pm$ 0.17 based on the low-resolution spectra of the LMC's YSO (targets included in the present samples). 
The present result is consistent with the CO$_2$/H$_2$O ratio of \citet{ST} within the uncertainties, but it is slightly lower than our previous result. 
It is mostly attributed to the effect of the PAH 3.3\,$\mu$m emission, which was not resolved and thus not correctly counted in the previous study.
\citet{Oli09} have estimated the column densities of the CO$_2$ ice for some common objects with the present targets based on the 15.2$\mu$m feature and also revised some of the column density estimates of \citet{ST}. 
The CO$_2$ ice column densities estimated in \citet{Oli09} are slightly lower than our estimate in Table \ref{tbl_NG_ice}, but generally consistent within the uncertainties. 
They also suggest that the CO$_2$ ice abundance of the LMC samples is higher than in massive YSOs in our Galaxy (CO$_2$/H$_2$O $\sim$ 0.32, Fig.~\ref{NG_H2O_CO2}).

The high CO$_2$/H$_2$O ratio in the LMC indicated by the previous studies is now confirmed with the increased number of samples. 
In addition, the uncertainties of the derived column densities decreased from those of \citet{ST} thanks to the higher S/N, the higher spectral resolution, and the improved treatment of the curve-of-growth method. 
It is also worth mentioning that a relatively high CO$_2$/H$_2$O ratio of 0.32 $\pm$ 0.02 is observed toward Galactic low-- and intermediate-- mass YSOs \citep{Pon08}. 
Their samples also show a large scatter in the CO$_2$ ice abundance as seen in the present samples. 
A similar trend in the CO$_2$ ice abundance between the LMC's massive YSOs and Galactic low-- and intermediate--mass YSOs may indicate some similarities of star formation conditions between these different samples, but a qualitative explanation is an issue for future investigation.

  The derived CO/H$_2$O ice ratio of the LMC's YSO ranges from $\sim$0.01 to 0.2 (Table \ref{tbl_NG_ice}), and this is similar to what is observed toward Galactic massive YSOs \citep[$\sim$0.02 -- 0.2,][]{Chi98,Gib04}. 
Also, the distribution range of the H$_2$O ice column densities of our samples is nearly comparable to that of the Galactic YSOs. 
Therefore, we can assume that there should be a different physical or chemical environment that selectively enhances the production of CO$_2$ ice in the LMC.

It is widely accepted that the CO$_2$ ice in the cold envelope of a YSO is formed by grain surface reactions since gas-phase reactions are not able to produce the observed abundance of the CO$_2$ ice, while molecules like CO are mainly formed by the gas-phase reaction \citep[e.g., ][]{Mil91}. 
However, the formation mechanism of the CO$_2$ ice still remains unclear, and it is one of the key topics in astrochemistry. 
Laboratory experiments indicate that the CO$_2$ ice is efficiently produced by the UV photon irradiation to H$_2$O-CO ice mixtures \citep[e.g., ][]{Ger96,Wat07}. 
It is reported by several authors that the LMC has an order-of-magnitude stronger UV radiation field than our Galaxy owing to its active star formation and its metal-poor environment \citep{Isr86, Lee07}. 
Thus the strong UV radiation field is able to help the efficient formation of the CO$_2$ ice in the LMC.

 On the other hand, an alternative mechanism of the CO$_2$ ice formation has been proposed to explain the detection of the abundant CO$_2$ ice toward quiescent molecular clouds, where energy sources of UV photon are not expected \citep{Whit98}. 
A theoretical study suggests that the CO$_2$ ice can also be formed through the diffusive surface chemistry without UV irradiation \citep{Ruf01}. 
This model suggests that a sufficient abundance of the CO$_2$ ice can be produced at relatively high dust temperatures, but this process is quite sensitive to the dust temperature and other assumed conditions. 
Several studies report that the interstellar dust temperature in the LMC is generally higher than in our Galaxy based on far-infrared to submillimeter observations of diffuse emission \citep[e.g., ][]{Agu03, Sak06}. 
In addition, \citet{vanL10,vanL10_b} derived the dust temperatures of YSOs in the LMC and Small Magellanic Cloud (SMC) and suggest that the dust temperatures become higher in more metal-poor environments of the SMC. 
Thus the dust temperatures of the present samples may be systematically higher than their Galactic counterparts, which are in more metal-rich environments. 
Therefore the high dust temperature in the LMC can also be a possible cause for the enhanced production of the CO$_2$ ice. 
From the discussion, we conclude that the general properties of the LMC's environments, the strong UV radiation field, and the high dust temperature are responsible for the observed high CO$_2$ ice abundance.

 One of the interstellar molecular features that likely develop from UV photolysis is the 4.62$\mu$m ``XCN'' feature, which is probably due to the OCN$^{-}$ \citep[e.g., ][]{Pen99,Ber00}. 
This feature provides a diagnostic of the local radiation field of objects \citep{Spo03}. 
In our galaxy, the XCN feature is generally very weak or absent toward YSOs and dark clouds, but is very strong toward a few high-mass YSOs and Galactic center objects \citep{Gib04}. 
Thus this feature could suggest that the LMC's strong UV radiation field is a dominant factor for the high CO$_2$ ice abundance in the LMC. 
Some of our samples show a hint of this XCN feature on the shorter side of the CO ice feature (ST6, 7, 10). 
There may also be an additional component seen in the spectrum of ST4 and ST8, but the peak wavelength seems to be slightly shorter than the XCN feature. 
No clear trends are seen in the CO$_2$/H$_2$O ratio of the objects with possible detection of the XCN. 
However, we cannot separate this feature from unresolved CO gas lines because the spectral resolution is not high enough, and the detection of XCN is still tentative, so that 
no definite conclusion can be drawn from the present spectra. 
Future observations should provide the needed resolution.


\subsection{Correlation of ice abundance with YSO properties}  
 As described in $\S$2, the present samples basically possess similar characteristics (high-mass Class I objects). 
Despite this, column densities of the H$_2$O, CO$_2$ and CO ices and their ratio show large variations. 
To investigate which physical condition is dominant in the formation and evolution of ices, we  compare the chemical conditions of ices and the properties of individual YSOs. 
It is probable that the radiation from the central star dominates the circumstellar environment of a massive YSO. 
If the formation of ices is controlled by the radiation from the star, there should be some parameters of YSOs that correlate well with the ice abundances. 

\begin{figure*}[!htb]
\begin{center}
\includegraphics[width=15cm, angle=0]{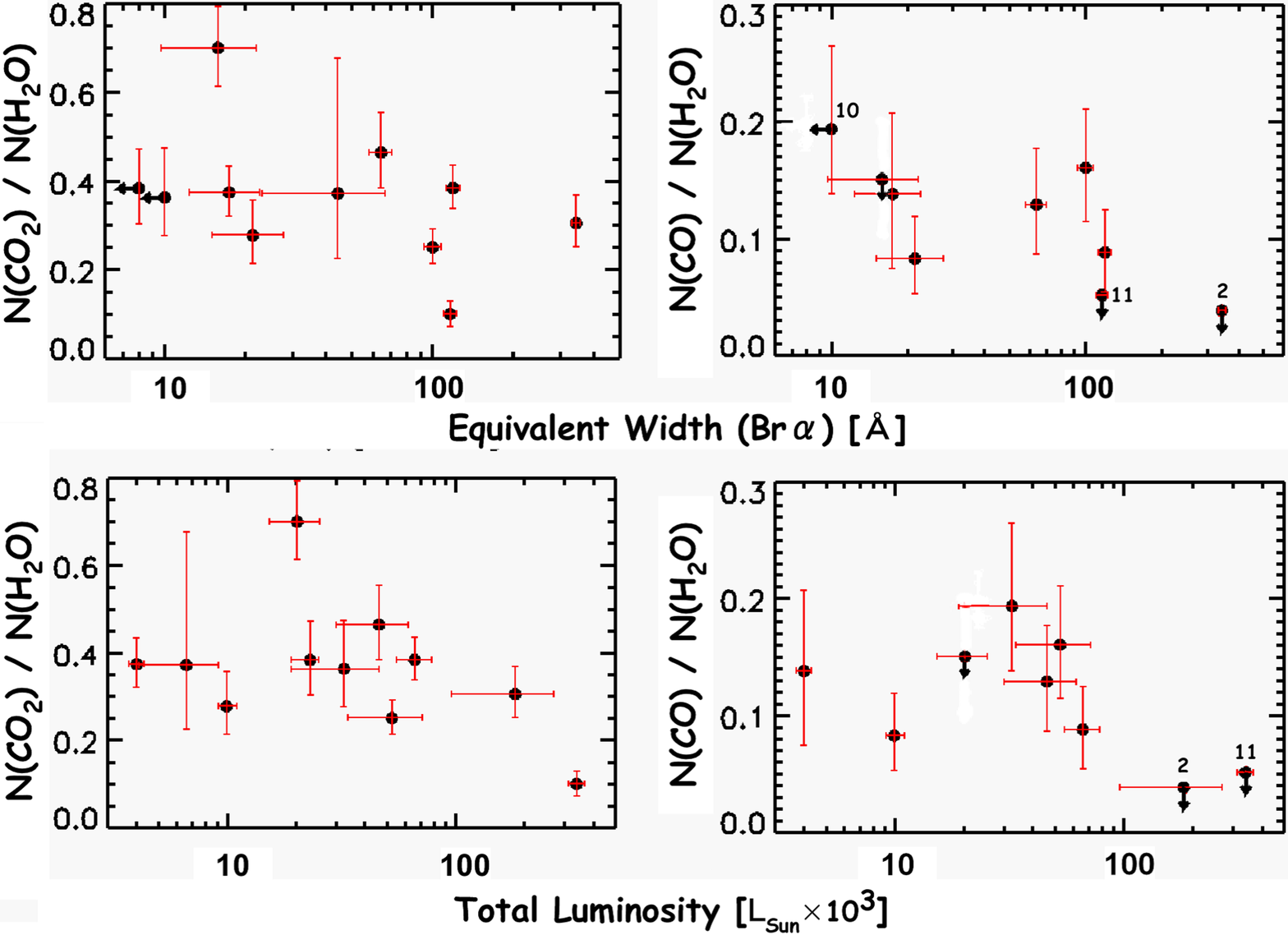}
\caption{The equivalent width of the hydrogen recombination line (Br$\alpha$) vs. CO$_2$/H$_2$O ratio (upper left) and CO/H$_2$O ratio (upper right). 
And a total luminosity of sample YSOs vs. CO$_2$/H$_2$O ratio (bottom left) and CO/H$_2$O ratio (bottom right). 
A YSO number given in Table \ref{tbl_NG} is plotted for the objects discussed in the text. 
A value of ST6 is not plotted due to its large uncertainties. 
). 
}
\end{center}
\label{corl}
\end{figure*}

 First, we compared the strength of the hydrogen emission lines with the CO$_2$ and CO ice abundance. 
Although it is not seen in the low-resolution spectra of \citet{ST}, the present spectra clearly detect hydrogen recombination lines toward the target YSOs. 
It is known that massive YSOs in their more evolved stage ionize their circumstellar gas by stellar UV radiation and form compact HII regions around the central star \citep{vanA00}. 
A variation in their hydrogen line strength thus indicates that the present targets are in their various evolutionary stages and possess different radiation environments (especially as for UV) in their circumstellar region, as discussed in \citet{Sea09}. 
We measured the equivalent width of the Br$\alpha$ line at 4.05$\mu$m, which is the strongest hydrogen recombination line in the wavelength range of our observations.

Figure \ref{corl} shows the comparison of the equivalent width of the Br$\alpha$ line with the CO$_2$/H$_2$O and CO/H$_2$O ice ratio. 
It seems that the abundance of the CO$_2$ ice does not correlate with the strength of the Br$\alpha$ line. 
The objects with similar CO$_2$ abundance show various Br$\alpha$ line strengths. 
This result indicates that the UV radiation from the central star is a less dominant factor for the evolution of the CO$_2$ ice. 
However, in the previous section, we mentioned that the UV irradiation can help form CO$_2$ ice. 
Several authors have demonstrated by simple calculation that the interstellar radiation field can penetrate the site where the CO$_2$ ice forms \citep[e.g.,][]{Whit98,Wat07}. 
We therefore suggest that the external (interstellar) UV radiation field may play an important role in CO$_2$ ice formation. 
On the other hand, it seems that objects with strong Br$\alpha$ line strengths (ST2, ST11) show a small CO ice abundance (upper limit), and one with the weakest Br$\alpha$ line (ST10) shows the highest CO ice abundance. 
Although the CO ice abundance of relatively weak Br$\alpha$ objects is rather scattered, the result indicates that the CO ice around YSOs is more likely to be affected by the stellar radiation than the CO$_2$ ice.

 A similar comparison was also made for the total luminosity of the YSOs, which is also a good estimate of the radiation environment of a YSO. 
The total luminosity was derived by the SED fitting described in $\S$2. 
As seen in Fig.~\ref{corl}, no clear correlation is seen between the CO$_2$ ice abundance and total luminosity of YSOs. 
The trend seen in the plot of the CO ice against the total luminosity is somewhat similar to what is seen in the comparison with the Br$\alpha$ strength; i.e., the overall correlation is weak, but the most luminous objects in the present sample (ST2 and ST11) show a small abundance of CO.

The low CO ice abundance toward luminous and probably evolved (indicated by strong Br$\alpha$ emission) YSOs can be explained by the low sublimation temperature of the CO ice. 
The sublimation temperature of the pure CO ice is lower than that of the pure H$_2$O and CO$_2$ ices \citep[16K, 50K, and 90K, for CO, CO$_2$, and H$_2$O,][]{Tie05}. 
Although the sublimation temperature changes depending on the chemical compositions of ices, their general trend does not change. 
The CO ice is therefore more easily affected by the stellar radiation owing to its low sublimation temperature. 
Given a similar geometry and distance between the central star and the circumstellar dust, the dust around more luminous YSOs is expected to be warmer since the radiation from the central star is the dominant heating source of their circumstellar dust. 
Likewise, dust around more evolved objects is expected to be warmer thanks to the dissipation of the circumstellar dust. 
The relatively high luminosity of ST2 and ST11 suggests that these objects might be clusters. 
However, we regard them here as YSOs of a similar kind as in other samples, because their infrared SEDs are well-fitted by the single YSO model of \citet{Rob07}, as well as in other samples. 
We thus suggest that the CO ice around the present luminous samples (ST2 and ST11) may be sublimated thanks to the intense radiation from the central star. 
The CO abundance of less luminous objects also has a scatter, which suggests that other factors than the stellar radiation may be important for the CO ice chemistry as for these objects.



\section{Summary}  
 To understand the chemical conditions of ices around YSOs in the Large Magellanic Cloud, we performed spectroscopic observations toward massive embedded YSOs and the candidates using the IRC onboard the infrared satellite {\it AKARI}. 
We obtained near-infrared spectra (R$\sim$80, 2.5--5um) of 12 objects in total. 
The ice absorption features of H$_2$O, CH$_3$OH, CO$_2$, $^{13}$CO$_2$, CO, and possibly XCN were detected in the spectra. 
In addition, hydrogen recombination lines and PAH emission bands were detected toward the majority of the targets.

Compared to our previous study \citep{ST}, the increased number of LMC's YSO samples, the higher spectral resolution, and the improved treatment of the curve-of-growth method make it possible to derive column densities of ices and their typical abundances more accurately. 
Also, the detection of hydrogen emission lines and the SED analysis of individual YSOs allow discussion of the correlation between ice abundances and YSO properties.

We derived the column densities of the H$_2$O, CO$_2$ and CO ices by fitting the laboratory ice spectra (for 3.05$\mu$m H$_2$O ice feature) and by the curve-of-growth method (for 4.27$\mu$m CO$_2$ and 4.67$\mu$m CO ice features).
As a result, it is shown that the typical CO$_2$/H$_2$O ice ratio of our samples ($\sim$0.36 $\pm$ 0.09) is higher than that of Galactic massive YSOs ($\sim$0.17), while the CO/H$_2$O ice ratio is comparable to Galactic ones. 
The systematical difference in the CO$_2$ ice abundance around the LMC's massive YSOs, which was suggested by previous studies \citep{ST,Oli09}, is now confirmed with the new near-infrared data. 
It is also shown that the scatter of the CO$_2$/H$_2$O ratio in the LMC is larger than for Galactic ones.

Although the formation mechanisms of the CO$_2$ ice is still unclear, we suggest that the strong ultraviolet radiation field and/or the high dust temperature in the LMC are responsible for the observed high abundance of the CO$_2$ ice. 
Our result provides evidence that YSOs in the metal-poor LMC maintain different chemical conditions in their circumstellar environments.

A relatively high CO$_2$ ice abundance ($\sim$0.32) and a large scatter are also reported toward Galactic low-- and intermediate-- mass YSOs. 
A similar trend of the CO$_2$ ice abundance between the LMC's massive YSOs and Galactic low-- and intermediate--mass YSOs may indicate some commonality in these objects, but a qualitative explanation considering both the structure of a YSO envelope and the ice chemistry is required in the future.

 The present YSO samples possess similar characteristics in their infrared SEDs, but they show a large variation in the CO$_2$ and CO ice abundances and in their near-infrared emission features. 
We compared the CO$_2$ and CO ice abundances with the strength of the hydrogen recombination line and total luminosity of the sample YSOs. 
It is shown that the CO$_2$ ice abundance and these YSO characteristics do not show a correlation. 
The present result indicates that the internal stellar radiation does not play an important role in the evolution of the CO$_2$ ice around a massive YSO. 
We thus need further investigations to shed light on the dominant physical factors in the CO$_2$ ice formation. 
Similar to the CO$_2$ ice, a clear correlation is not seen between the CO ice abundance and the YSO properties. 
However, it is suggested that the CO ice abundance of luminous and probably evolved samples are significantly lower than other samples. 
We infer that more volatile molecules like CO are susceptible to the effect of the stellar radiation. 

 The present results revealed the presence of a massive YSO that shows a CO$_2$/H$_2$O ratio of up to $\sim$0.7. 
If there is also a chemical difference in the LMC's lower mass YSOs, it will indirectly tell us the chemical diversity of subsequently-formed planets since the interstellar ices are thought to be an origin of planetary system ices, as discussed in \citet{Ehr00}. 
The future infrared telescope missions SPICA, JWST, and SOFIA will allow better understanding of the diversity of solid materials in the universe.


\begin{acknowledgements}
 This work is based on observations with {\it AKARI}, a JAXA project with the participation of ESA. 
We are deeply grateful to the referee J. Th. van Loon for his useful comments, which greatly improved this paper.
We thank all the members of the {\it AKARI} project for their continuous help and support. 
We are grateful to Youichi Ohyama for kind support in the analysis of spectroscopic data. 
We also thank Helen Fraser and Jennifer Noble for fruitful discussions in Glasgow. 
This research has made extensive use of the Leiden University's Laboratory ice database. 
This work is supported by a Grant-in-Aid for Scientific Research from the JSPS (No. 180204014).
\end{acknowledgements}




\appendix
\section{The curve-of-growth method}  
We calculated column densities of unresolved ice features by using the equivalent width. 
This calculation technique is called the ``curve-of-growth'' method, and it is traditionally used to study gaseous absorption lines \citep{ToMS}. 
Since the optical depth is the logarithm conversion of the normalized flux (1 - F$_{obs}$/F$_c$), the peak optical depth derived from unresolved features is not a true value. 
Thus, in case of our low-resolution spectra, we will significantly underestimate a column density if it is derived from the simple integration of the optical depth as expressed in Eq. (1), and this is why we need to use the curve-of-growth method. 
This method was applied for the first time to absorption features of ices in \citet{ST}, but we overestimate the uncertainties and a detailed description of the method was not given because of space limitations. 
In this appendix, we describe our improved treatment of the curve-of-growth method. 

\begin{table}
\caption{Ice profiles used for calculating the curve-of-growth}
\label{CoG_ice}
\centering
\renewcommand{\footnoterule}{}  
\begin{tabular}{ c c }
\hline\hline                                                               \\
Molecule       & Ice Profile                                       \\
\hline \\
CO$_2$         & H$_2$O : CO$_2$ ice mixture at 14K                   \\
CO             & Gaussian profile (FWHM=6.89 cm$^{-1}$, $\lambda_{center}$=4.67$\mu$m)   \\
\hline                                                                     \\
\end{tabular}
\end{table}

\begin{figure*}[!t]
\begin{center}
\includegraphics[width=14cm, angle=0]{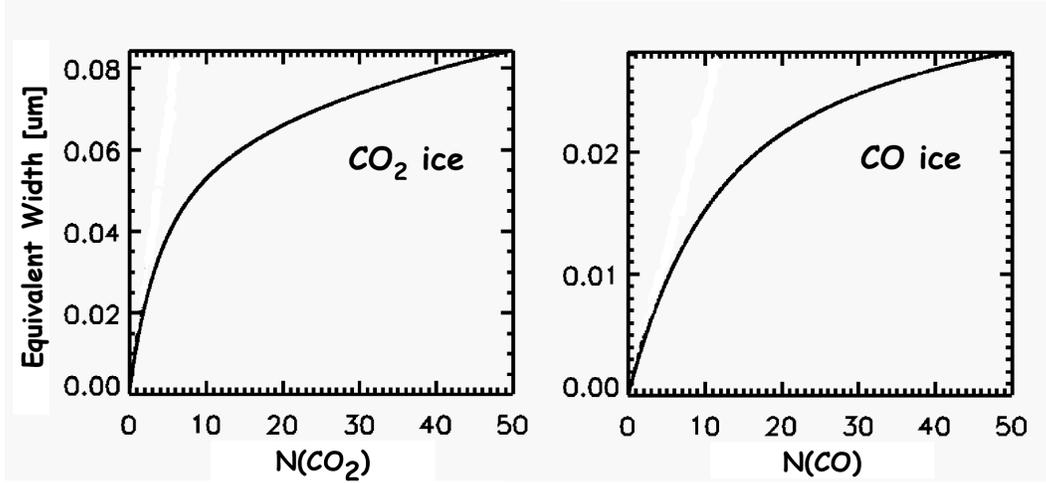}
\caption{Calculated curve-of-growth for the CO$_2$ ice (left) and the CO ice (right). 
These plots are used to convert measured equivalent widths to column densities. 
The uncertainties of derived column densities increased for deeper absorptions. 
}
\label{CoG}
\end{center}
\end{figure*}

 First, we convert the measured equivalent width into a column density. 
To do this, we need to assume intrinsic absorption profiles of target ice features. 
Here, they are the 4.27$\mu$m CO$_2$ ice and 4.67$\mu$m CO ice feature. 
For CO$_2$, we considered the typical compositions and temperature of the interstellar CO$_2$ ice and chose an appropriate laboratory spectrum as shown in Table \ref{CoG_ice} \citep[data taken from the Leiden Molecular Astrophysics database]{Ehr96}. 
For CO, it is reported that several combinations of laboratory ice profiles can account for the same observed spectrum and the fitting degenerates \citep{Pon03}. 
We therefore used a single Gaussian profile that is close to the typically observed CO ice feature around massive YSOs \citep{Chi98}. 
The assumed profiles of the CO$_2$ and CO ices features are summarized in Table A.1. 
By using these profiles, we calculated both equivalent widths and column densities for the various optical depths. 
The plot of the derived equivalent widths vs. column densities is shown in Fig.~\ref{CoG}, and this is called the curve-of-growth. 
The equivalent widths can be measured from observations, and these plots are used to derive column densities of ices;
however, the uncertainty becomes larger for the higher column densities owing to the saturation effect of the curve-of-growth.

\begin{figure*}[!]
\begin{center}
\includegraphics[width=14cm, angle=0]{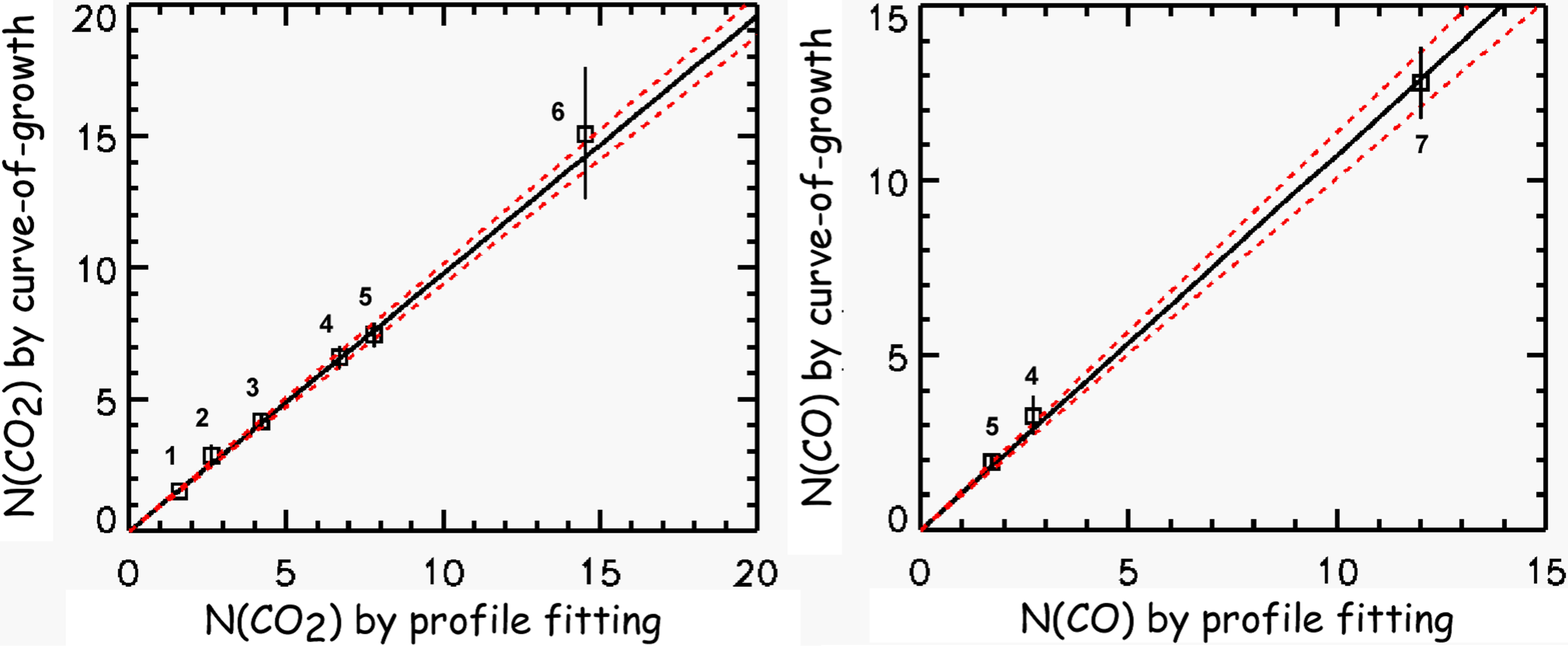}
\caption{Comparison of the CO$_2$ and CO ice column densities derived from the low-resolution spectra by the curve-of-growth and those derived from the high-resolution {\it ISO}-SWS spectra by the laboratory ice fitting \citep{Ger95, Gib04}. 
The axes are expressed in units of 10$^{17}$ cm$^{-2}$. 
Solid and dashed lines indicate the result of the linear fit and its one sigma error. 
The labeled numbers in the figure correspond to the name of the objects that are used in this analysis as follows; 1. Mon R2 IRS3, 2. Orion IRC2, 3. S140, 4. Elias 29, 5. AFGL 2136, 6. W33A, 7. NGC7538 IRS9. }
\label{CoG_err}
\end{center}
\end{figure*}
 It is known that the actual profile of the observed ices in star-forming regions is not uniform since they vary depending on the compositions and temperature of ices \citep{Ger99,Gib04}. 
The derived column densities depend on the assumed profiles, so 
we need to check the validity of our method and to evaluate the uncertainties caused by the method. 

 We first prepared 2--5$\mu$m {\it ISO}-SWS spectra of embedded massive YSOs\footnote{Data taken from http://isc.astro.cornell.edu/~sloan/library/swsatlas/atlas.html} \citep{Slo03}, whose ice column densities are already presented in \citet{Gib04}. 
The spectral resolution of these data is high enough to resolve narrow ice features (R $\sim$500--1000). 
Then their spectral resolution was degraded to the resolution of the {\it AKARI} IRC NG mode (R $\sim$80) by convolving the point spread function of the IRC. 
Next, we derived column densities of the CO$_2$ and CO ice by these converted spectra using the same method as that described in $\S$4.2.2. 
The derived column densities and the results of \citet{Gib04} are plotted in Fig.~\ref{CoG_err}. 


The figure indicates that the column densities derived from our curve-of-growth method are basically in good agreement with the results of \citet{Gib04}, which derived the column densities from the high-resolution spectra. 
But a systematic difference and scatter are also seen in the figure. 
These are mainly caused by the difference between the observed and assumed ice profiles. 
The presence of weak emission or absorption features may also be responsible for these differences. 
We obtained the following equations from a straight-line fit to the data points in Fig.~\ref{CoG_err}. 
\begin{equation}
N(CO_2) = N_{G04} \times (0.98 \pm 0.038)
\end{equation}
and
\begin{equation}
N(CO)   = N_{G04} \times (1.08 \pm 0.065), 
\end{equation}
where $N_{G04}$ represents the column density of each ice presented in \citet{Gib04}. 
In \citet{ST}, we adopted 20$\%$ uncertainties to account for the above systematic difference in the derived column density of the CO$_2$ ice and added to the observational uncertainties. 
The uncertainties were, however, overestimated as expressed in Eq. (A.1). 
In this study, the column densities derived from the curve-of-growth method are divided by the slope of the corresponding line to correct the systematic difference in the derived column densities. 
In addition, a 1-$\sigma$ fitting error of the slope is considered as an uncertainty of our curve-of-growth method and added to the uncertainties in the observation and the fitting.

\end{document}